\documentclass[conference]{IEEEtran}
\IEEEoverridecommandlockouts
\usepackage{cite}
\usepackage{amsmath,amssymb,amsfonts}
\usepackage{algorithmic}
\usepackage{graphicx}
\usepackage{textcomp}
\usepackage{xcolor}
\def\BibTeX{{\rm B\kern-.05em{\sc i\kern-.025em b}\kern-.08em
    T\kern-.1667em\lower.7ex\hbox{E}\kern-.125emX}}
\begin{document}

\title{Load Estimation for Industrial Load-lifting Exoskeletons Using Insole Pressure Sensors and Machine Learning\\

\thanks{The authors would like to thank Fei Zhang for assistance with experimental data collection. The code and dataset used in this study are publicly available at: {https://github.com/DecemberDawn-Gate/insoe-lifting-load-estimation} }
}

\author{
\IEEEauthorblockN{Kaida Wu, Peihao Xiang, Chaohao Lin, Ou Bai*}
\IEEEauthorblockA{\textit{Department of Electrical and Computer Engineering} \\
\textit{Florida International University}, Miami, US \\
 *Corresponding author’s e-mail: kwu020@fiu.edu, obai@fiu.edu}
}

\maketitle

\begin{abstract}

To enhance lifting-load estimation accuracy in industrial upper-limb assistive exoskeletons, this study proposes a machine learning-based approach using insole pressure sensors. Unlike traditional methods that rely on electromyography (EMG), force sensors, or posture data, insole pressure sensors provide a non-invasive, posture-independent, and stable solution suitable for long-term use. Lifting load data ranging from 2 to 10 kg (0.5 kg intervals) were collected from five subjects. Two data representations were investigated: channel-based vectors and map-based images. For the channel-based approach, conventional regression models (SVR, MLP, and Elastic Net) were trained on pooled data from all subjects to assess inter-subject generalization, specifically testing the ability to infer load levels unseen during training. In parallel, a preliminary feasibility study was conducted for the map-based deep learning model (MobileNetV2) using inner-subject data. Results indicate that the channel-based SVR achieved the most balanced accuracy and generalization performance, with a mean absolute error of 0.547 kg. These findings demonstrate the feasibility and advantages of using insole pressure data for variable load estimation, supporting control strategies in industrial exoskeleton applications.

\end{abstract}


\section{Introduction}

Load-lifting exoskeletons are able to help minimize fatigue, reduce injuries, and enhance workforce productivity \cite{kuber2022systematic}. Accurately estimating the load held by the wearer is a necessary condition for an effective control strategy in load-lifting exoskeletons \cite{nasiri2022human}. 
\vspace{0em}
When performing tasks such as picking up, carrying, and placing heavy objects, the human body typically exhibits slow dynamic characteristics. As a result, gravitational torque accounts for a significant portion of the total joint torque required. Compensating for gravitational torque can effectively reduce the muscular effort of the user \cite{nasiri2022human}\cite{khan2015adaptive}. Accurately estimating the  load held by the user is critical to prevent overcompensation (excessive assistance, loss of control) and under-compensation (insufficient support, increased fatigue) in exoskeleton applications – especially critical for unknown or variant load; for lifting tasks, however, the load is variant and needs real-time estimation. For dynamic control in exoskeleton applications, which are mainly for rehabilitation purposes, the focus is on dynamic compensation while they assume the load, i.e., body weight is a constant without change so that the constant gravity compensation for the weights can be previously given and optimized \cite{gull2020review}. For the adaptive impedance control, loads should be accurately estimated (or be given) in order to provide appropriate stiffness \cite{aguirre2007active}. The objective of this study is to investigate efficient methods to accurately estimate varying loads for exoskeleton applications. 


\section{Related Work}

\indent Current load estimation methods for upper-limb exoskeletons fall into three main categories: force/torque sensors, muscle contraction torque estimation, and posture-based estimation. In master-slave augmentation exoskeletons like those by Lee \cite{lee2014human}, load is measured directly via force/torque sensors at the end-effector. However, this type of exoskeleton is controlled by the handle, with the full load borne by the exoskeleton rather than the human body, which compromises the user's natural grasping ability. 

Lifting load can be estimated from muscle contraction indirectly. Muscle contraction torque estimation typically relies on bio-signals or muscle deformation, often combined with muscle models. Several studies have explored different approaches for load estimation. Totah \cite{totah2018low} and Aziz \cite{aziz2019electromyography} employed surface electromyography (sEMG) signals for load classification. Totah's method achieved 80\% (±10\%) to 81\% (±7\%) accuracy for loads of 0, 10, and 24 lbs, while Aziz's SVM model reached 99\% accuracy for 1 kg, 3 kg, and 7 kg. For muscle deformation, Kim \cite{kim2014development} utilized muscle circumference sensors to estimate elbow torque for loads of 5 Nm, 10 Nm, and 15 Nm, with errors of 25\%, 24\%, and 22\%. Islam \cite{islam2019payload} applied force myography (FMG) for loads of 0.8 kg, 2.5 kg, and 4 kg, achieving mean absolute errors between 0.14 kg to 0.37 kg. Although these methods demonstrate promising accuracy, they are sensitive to external interference and individual physiological differences, which can compromise long-term reliability. 

Another approach for lifting load estimation is based on body postures.  Pesenti \cite{pesenti2023imu} employed inertial measurement units (IMUs) to classify 5 kg, 10 kg, and 15 kg loads, achieving classification accuracies of 81.99\%, 88.16\%, and 91.7\%, respectively. While IMU-based methods are non-intrusive and easy to wear, their reliability is highly dependent on consistent posture, limiting their applicability in dynamic lifting tasks.

These existing approaches highlight the challenges in achieving robust and generalizable load estimation. The successful application of insole sensors for human body-weight estimation inspired our exploration of insole sensor-based estimation as a more stable and posture-independent alternative. By leveraging plantar pressure redistribution induced by external loads, the proposed approach provides a non-invasive, posture-independent, and stable alternative to conventional bio-signal–based and posture-based load estimation methods.

The potential advantages of insole sensor-based load estimation approach for exoskeleton applications:

\begin{itemize}
\item Ease of Integration: Insole sensors are non-invasive, easy to wear, and do not require complex electrode placements or calibration, making them a practical choice for real-world applications.
\item Reduced Sensitivity to External Interference: Unlike sEMG and force myography, which are prone to signal variations due to sweat, skin impedance, or sensor placement, insole sensors provide a more stable measurement.
\item Improved Long-term Reliability: Since insole sensors measure force distribution through the feet, they are less susceptible to individual physiological differences (e.g., muscle fatigue, skin condition) that can affect bio-signal-based methods.

\item Posture Independence: Unlike IMU-based methods, which require consistent body posture, insole sensors can estimate load variations without being affected by upper body movements.
\end{itemize}

We propose to employ insole sensors and state-of-the-art machine-learning technology for accurate varying-load estimation in upper limb exoskeletons specifically addressing load lifting tasks. The contributions of this study are as follows:
 
\begin{itemize}
\item A novel method for accurate varying load estimation by using insole sensors.
\item Under channel-based representation, we evaluated three regression methods (SVR, MLP, Elastic Net) for load estimation and their generalization to unseen loads, offering algorithmic guidance for modeling and deployment of pressure sensor data.
\item A preliminary evaluation was conducted under the map-based representation by converting insole pressure data into image inputs, exploring the performance differences between full fine-tuning and linear probing on a pretrained MobileNetV2. This provides a feasibility validation and strategic reference for thermal map-based pressure data modeling.
\end{itemize}
\vspace{1em} 
\section{Methodology}

\subsection{Implementation and adaptation of insole sensors}

We employed a cost-effective insole pressure sensor (RX-ES42-18, Guantuo Electronic Technology Co., Guangdong, China, priced at \$100 per suit). The EU size 42 insoles are 260 mm long with 36 channels (18 per insole) in a distributed array, each capable of a maximum load of 70 kg. Powered by a 3.7 V battery, the device transmits data via Bluetooth; each data set includes a timestamp and is processed internally before being recorded on the host computer at a 20 Hz sampling rate. For our experiment, we implemented a rationalized deployment. 

\begin{figure}[htbp]

	\centering
	\includegraphics[width=3.2in]{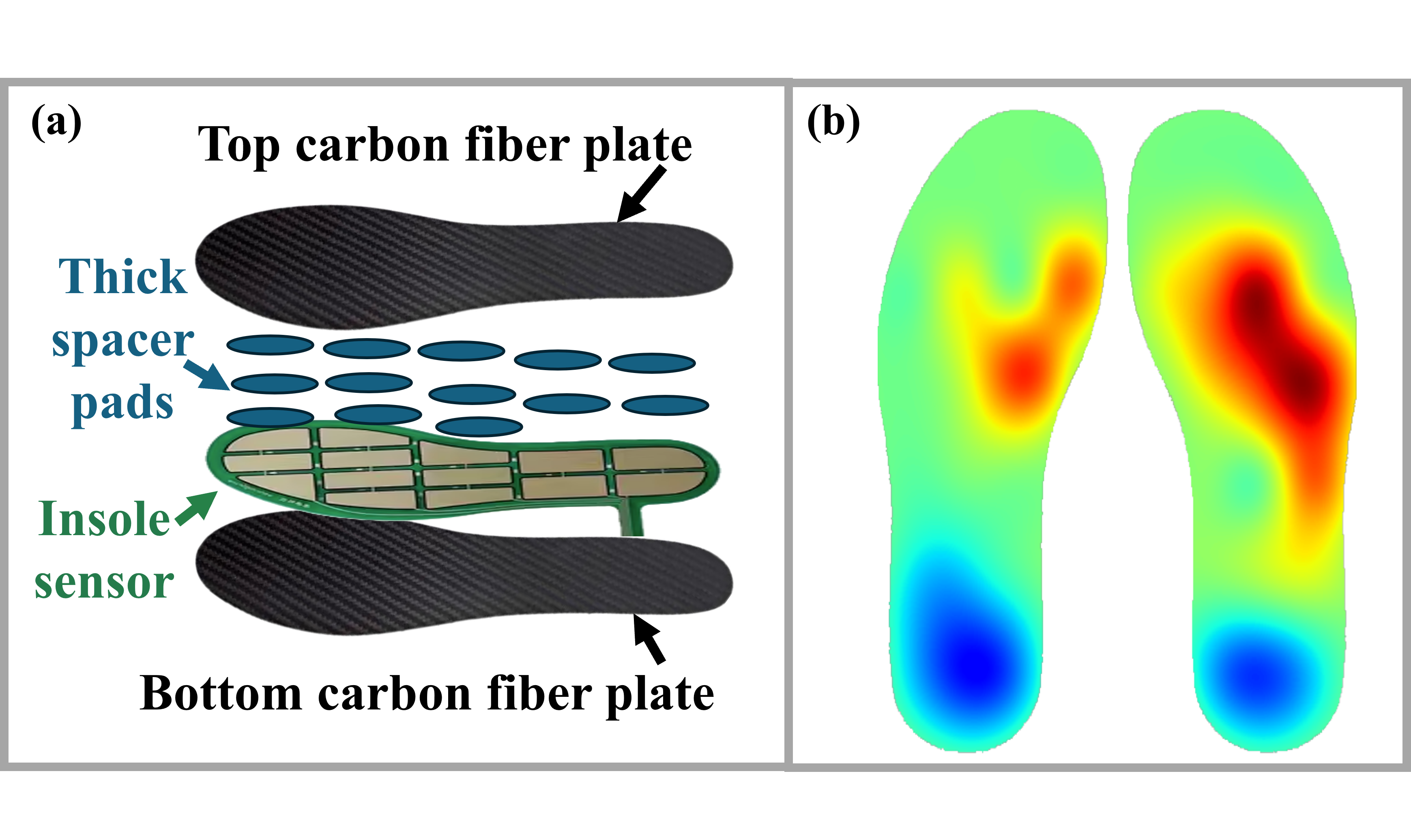}
	\caption{(a) The overall layers of adapted insole sensor. (b) The insole thermal-map of a real pressure data}
	\label{Figure1}
    
\end{figure}

To minimize the shoe sole's influence, we adopted a pancake structure \cite{muzaffar2020shoe}. The insole sensor was fixed onto a rigid carbon fiber plate shaped to the foot. We then 3D-printed 2 mm-thick spacer pads for each measurement channel to enhance foot-sensor contact, and placed another rigid carbon fiber plate over them for stability. The setup is shown in Fig.\ref{Figure1} (a), and the adapted sensor was placed inside a pair of experimental shoes.

\subsection{The data collection and experiment description }

As this study is in its early stages, we aim to explore the feasibility and effectiveness of using insole sensors for load estimation. The experiment collected data from five subjects with different body weights ($73.12 \pm 11.04$ kg) and foot sizes (eu $39 \pm 3$). All subjects were in good health, with no known illnesses or injuries. Prior to participation, they were fully informed about the experimental details and voluntarily consented to the study and the use of their data for research and publication. The experimental procedures were reviewed and supervised by the Institutional Review Board (IRB).

\begin{figure}[htbp]
	\centering
	\includegraphics[width=3.3in]{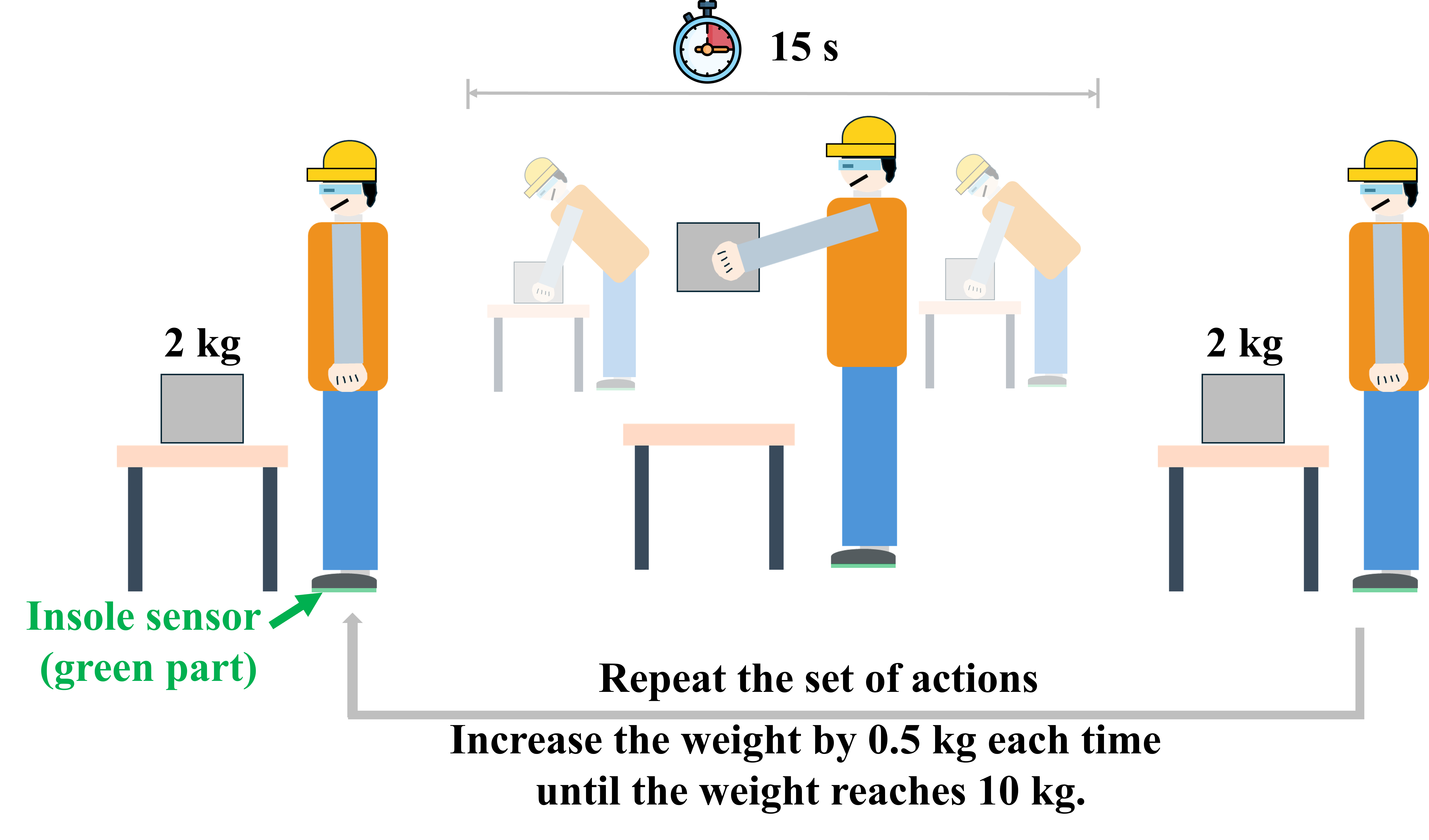}
	\caption{Insole sensor data collection from 2 kg to 10 kg loads}
	\label{Figure2}
\end{figure}

Each subject completed three experimental sessions on separate days to minimize fatigue and ensure comfort. Loads ranged from 2 kg to 10 kg using dumbbell weights. During each session, subjects wore experimental shoes and stood or lifted the load in a comfortable posture, using a container box placed on a table. A computer served as a timer and collected insole sensor data via Bluetooth. Every 15 seconds, a beep signaled the subject to switch actions: starting from standing still, lifting and holding the box, then returning it to the table. After each cycle, 0.5 kg was added until the load reached 10 kg, ending the session. This collection is shown in Fig. \ref{Figure2}.
\subsection{Data Preprocessing}
To evaluate the accuracy and generalization ability of regression models based on insole pressure sensor data for lifting-load estimation, we designed an assessment protocol that accounts for inter-subject variability and tests the model's ability to predict unseen loads—load levels not labeled during training. Specifically, for each of the five subjects, the first two rounds of data were used for training, while the third round served as the test set to assess regression performance. Additionally, to further examine the model's extrapolation capability to unseen load levels, we deliberately excluded all samples corresponding to 3 kg, 6 kg, and 9 kg during training, retaining them only in the test set. This setup enables evaluation of the model's generalization to previously unseen loads.

This experiment uses channel-based and map-based methods for model training. The channel-based approach treats the 36 channels as a one-dimensional feature vector, while the map-based approach converts them into a two-dimensional foot pressure thermal-map. Only spatial information is used, with each sample considered independent, ignoring temporal information.

After data collection, the 36-channel insole data from each session are segmented by timer timestamps and normalized. A second-order Butterworth low-pass filter with a 0.3 Hz cutoff frequency is applied to retain low-frequency pressure features for model training. Since industrial lifting tasks are dominated by quasi-static or low-frequency dynamics, higher-frequency components primarily reflect noise or transient motion rather than load magnitude. The filtered data are divided into baseline (no load) and load-lifting periods using 15-second intervals, with the middle 5 seconds extracted for baseline and 10 seconds for load-lifting to ensure stability and sufficient data.

To reduce variations from body weight or shoe fitting, raw pressure values are replaced with differential values. For each load-lifting segment, the average pressure from the preceding baseline is subtracted from each channel's data, making the differential values directly correspond to actual loads. The baseline value was obtained from the insole pressure sensor during no-load lifting conditions.

For map-based model training, we interpolated 36-channel data into a 2D plantar pressure thermal-map. A custom program was used to mark the contour coordinates of each channel. The differential values were normalized, with the color scale ranging from deep blue (minimum) to deep red (maximum), based on the maximum and minimum within $±2 \sigma $ (2 standard deviations) from the mean across subjects. The thermal map was smoothed, cropped, and resized to 224*224 pixels for model input. Fig. 1 (b) shows the thermal map of subject 1 under an 8.5 kg load. Map-based results are obtained only from subject 1 data and require further validation across subjects.

As this is an initial exploration, the map-based method was trained using data from subject 1 only, and experiments were conducted solely in an inner-subject setting. For fair comparison, SVR was also trained using only subject 1's data in this context. In contrast, for the channel-based approach, models were trained on data from all subjects and evaluated in inter-subject settings to assess generalization ability.

\subsection{Methods for load estimation}
We use three machine learning models-Elastic Net, Support Vector Regression (SVR), and Multi-Layer Perceptron (MLP)-to estimate lifting loads between 2-10 kg. Elastic Net \cite{ogutu2012genomic} is a linear regression method combining L1 (Lasso) and L2 (Ridge) regularization to minimize mean squared error. SVR \cite{awad2015support} applies kernel functions to map inputs into high-dimensional spaces, capturing nonlinear relationships. MLP \cite{taud2017multilayer}, with multiple hidden layers, effectively models complex nonlinear patterns .

\indent We adopt MobileNetV2 \cite{sandler2018mobilenetv2}, a lightweight convolutional neural network designed for efficient image tasks, to process two-dimensional pressure thermal maps. Two transfer learning strategies are used: Fully Fine-Tuning, which updates the entire network (optionally with added MLP layers), and Linear Probing, which freezes MobileNetV2 and trains only added layers.

\indent Given the sensor sampling interval of 50 ms, we designed the model to produce one output every 500 ms. Although the regression model operates on individual samples, a temporal aggregation strategy is applied at the output stage to enhance stability. Specifically, the model performs estimation at each sensor timestep, and after accumulating 10 predictions, the final output is computed as the mean of these predictions using Quantile-based trimming, which reduces the influence of outliers by discarding predictions outside predefined quantile bounds.

\indent During model training, the first and second sessions of data from all five subjects were used as the training set, and the third session was reserved for testing. Instead of training individual models per subject, data from all subjects were pooled to enhance generalization. The training data were further split into 80\% for training and 20\% for validation. For each model, three different hyperparameter configurations were tested to determine the optimal setup, with cross-validation applied during training.

For evaluation and comparison of model performance, we used Mean Absolute Error (MAE) as the metric. To assess the statistical significance of the performance differences between the models, we employed the Mann-Whitney U test. A p-value of less than 0.05 was considered statistically significant.

\subsection{Experimental Hyper-parameter Settings}
In the channel-based model, SVR uses a polynomial kernel (degree=2, C=1, gamma=1e-8, coef0=10, epsilon=1.3) to balance robustness and training error. The MLP model accepts 36 features, with hidden layers of 32, 16, and 8 neurons using ReLU, Batch Normalization after the first two layers, and Dropout (0.3). It is optimized with AdamW (lr=1e-4) using MSE loss, while Elastic Net is set with alpha=0.1 and L1 ratio=0.1. In the map-based model, the Fully Fine-Tuning approach unfreezes all MobileNetV2 parameters and replaces its head with two MLP layers (32 and 16 neurons, ReLU, Dropout 0.5), whereas Linear Probing keeps MobileNetV2 frozen and adds the same MLP. Both methods use AdamW (lr=1e-4) and MSE, ensuring consistent hyper-parameters for fair model comparisons.

\vspace{1em}
\section{Results and Discussion}
\subsection{Results}
In the channel-based approach, the training dataset contains 27,300 samples with 36 features, split into 80\% training (21,840 samples) and 20\% validation (5,460 samples), while the test set has 1,407 samples. For the map-based approach, the training and test datasets have shapes of (10,268, 224, 224, 3) and (5,134, 224, 224, 3), respectively.

\begin{figure*}[htbp]
	\centering
	\includegraphics[width=6in]{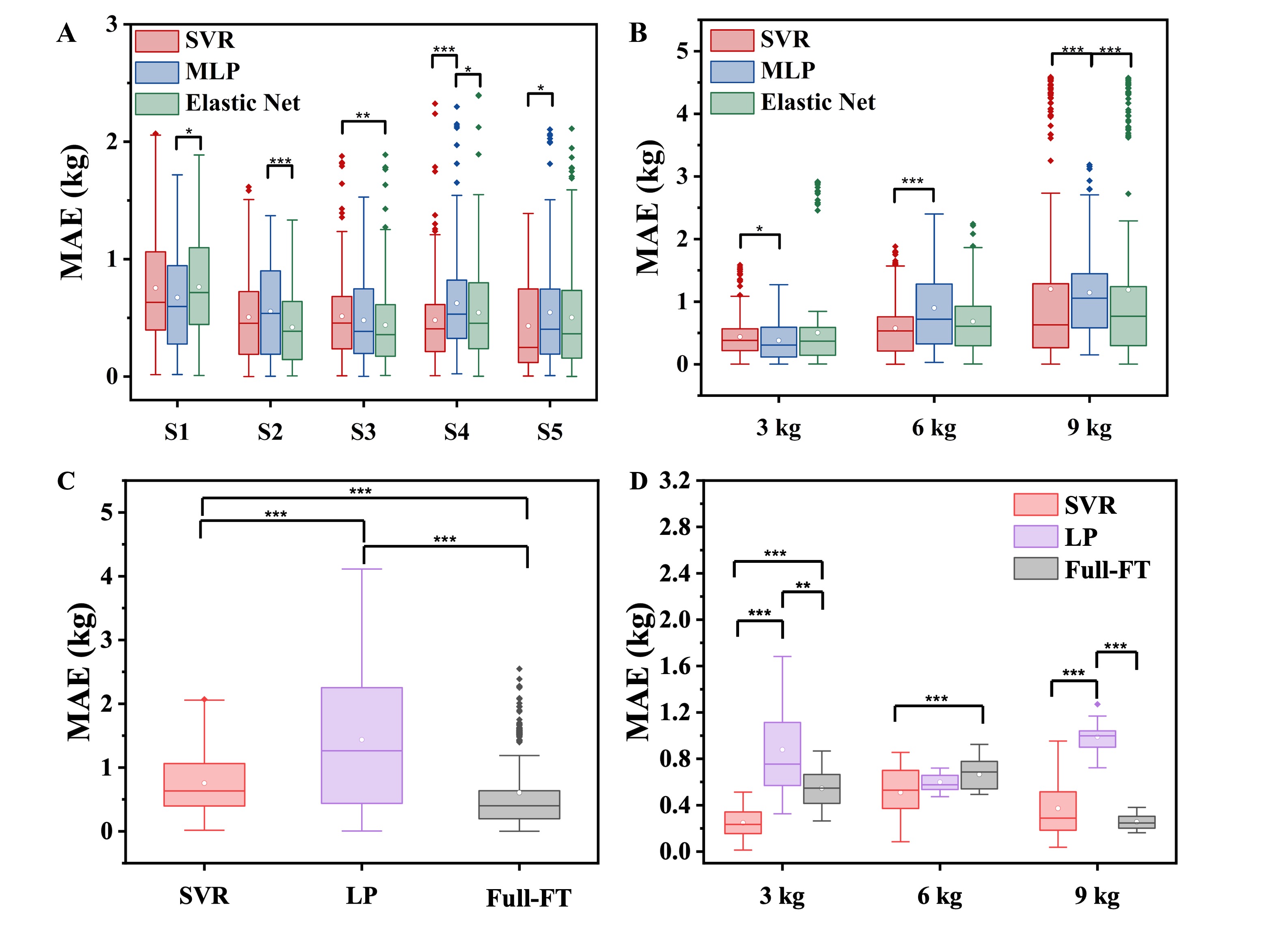}
	\caption{(A) MAE distribution of Channel-based approaches, which are SVR, MLP, and Elastic Net, across five subjects (subject 1 to 5) using insole pressure data. The load range is 2 kg to 10 kg. (B) MAE distribution for unseen load levels (3 kg, 6 kg, 9 kg) to evaluate models inference capability . (C) Performance comparison of channel-based approach ( SVR ) and map-based approach ( MobileNetV2 with Linear Probing (LP) and Full Fine-Tuning (Full-FT))on load estimation. (D) Under the map-based approach, MAE distribution for unseen load levels (3 kg, 6 kg, 9 kg) to evaluate models inference capability.}
	\label{Figure3}
\end{figure*}

The predictive performance of the three models (SVR, MLP, and Elastic Net) across five subjects (S1–S5) within the 2 to 10 kg loads is illustrated in Figure 3(A). The experimental results indicate that model efficiency varies significantly across individuals. For subject S1, MLP achieved a slightly lower median MAE compared to the other models, showing a significant difference when compared with Elastic Net ($p < 0.05$). However, in the cases of subjects S2 and S3, Elastic Net demonstrated superior performance, maintaining lower median and mean MAE values. Specifically, Elastic Net significantly outperformed MLP for S2 ($p < 0.001$) and SVR for S3 ($p < 0.01$). For subjects S4 and S5, SVR exhibited a clear predictive advantage. Statistical analysis revealed that in S4, the MAE of SVR was significantly lower than both MLP ($p < 0.001$) and Elastic Net ($p < 0.05$). Similarly, SVR maintained the lowest error distribution for S5, with a significant difference observed against MLP ($p < 0.05$). Regarding the stability of predictions, MLP produced a higher density of outliers in S4 and S5, with maximum MAE values approaching $2.5$ kg, indicating greater volatility in these specific subjects. In contrast, SVR (mean MAE = 0.547 kg) and Elastic Net (mean MAE = 0.536 kg) displayed more compact box structures across most subjects.

The Figure 3(B) illustrates the distribution of MAE for the three regression methods under different unseen load conditions (3 kg, 6 kg, and 9 kg), based on insole pressure sensor inputs. This evaluation aims to assess the models' inference and generalization capabilities for unseen loads. Overall, the MAE for all models exhibited an upward trend as the load weight increased, with the largest errors and most significant variance observed at the 9 kg level.

Under the 3 kg load, all models maintained relatively low error levels, with medians below 0.5 kg. A significant difference was observed between SVR and MLP ($p < 0.05$), where SVR showed a slightly more concentrated error distribution despite some low-level outliers. For the 6 kg load, the performance gap between models became more pronounced. SVR achieved the highest precision, with its MAE significantly lower than that of MLP ($p < 0.001$). Elastic Net performed moderately, staying between SVR and MLP in terms of median error and distribution spread. At the 9 kg load, the prediction task became considerably more challenging. While the median MAE for all models rose to approximately 0.7–1.1 kg, SVR and Elastic Net significantly outperformed MLP ($p < 0.001$). Notably, SVR and Elastic Net exhibited a high density of upper outliers reaching up to 4.5 kg, suggesting that while they maintain better central tendency than MLP at high loads, they are still susceptible to extreme prediction errors under these conditions. In contrast, although MLP had fewer extreme outliers at 9 kg, its overall box position (interquartile range) was notably higher than that of the other two models.

Figure 3(C) shows the MAE distribution for SVR, MobileNetV2 with Linear Probing (LP), and Full Fine-Tuning (Full-FT) on the load estimation task on subject 1. Full-FT achieved the best performance (MAE = 0.605 kg), with the lowest median error and compact distribution, despite some low-range outliers. SVR followed with stable performance. LP performed worst, showing high median error, wide spread, and greater variability, indicating poor and inconsistent predictions.

Figure 3(D) illustrates the distribution of MAE for SVR, LP, and Full-FT evaluated on subject 1 under three specific unseen load conditions (3 kg, 6 kg, and 9 kg). At the low load level (3 kg), SVR demonstrated the best performance, yielding an MAE significantly lower than both LP and Full-FT ($p < 0.001$). In this phase, Full-FT exhibited a relatively wider error distribution with a higher mean compared to SVR, while LP performed the worst, showing the largest error fluctuation. At the medium load level (6 kg), SVR maintained a low error level, though its advantage diminished compared to the 3 kg condition. Although SVR remained statistically superior to Full-FT ($p < 0.001$), the gap between the two narrowed, as evidenced by the closer medians and interquartile ranges in the box plots. LP showed some performance improvement in this stage but still lagged behind SVR. Crucially, a significant performance reversal was observed at the high load level (9 kg). The Full-FT demonstrated exceptional stability and precision, characterized by a highly compact MAE distribution that was significantly lower than both SVR and LP ($p < 0.001$). Conversely, as the load increased to 9 kg, the error range for SVR expanded noticeably (indicating increased variance), suggesting a decline in prediction stability under high loads. LP performed the poorest under high-load conditions, with error values diverging substantially.

 \subsection{Discussion}

The experimental results demonstrate the feasibility and effectiveness of using insole pressure sensors for estimating variable lifting loads in industrial upper-limb exoskeleton applications. Across five subjects and a wide load range from 2 to 10 kg, the proposed approach achieved stable estimation performance and maintained reasonable accuracy even under unseen load conditions. These findings indicate that plantar pressure redistribution induced by external loads provides sufficient information for load inference, supporting the use of insole sensors as a non-invasive and posture-independent alternative to traditional bio-signal–based or posture-based estimation methods. Crucially, the use of differential pressure values—subtracting the no-load baseline from active lifting data—mitigates the interference of individual body weight. This preprocessing strategy allows the models to focus on load-induced pressure increments, providing a solid physiological foundation for the observed cross-subject generalization. Although the subject pool is limited, the consistency of trends observed across all five participants, together with the unseen-load evaluation protocol, suggests that the learned models capture load-dependent plantar pressure patterns rather than subject-specific artifacts.

The comparative analysis of channel-based regression models reveals that simpler models, particularly SVR and Elastic Net, exhibit superior robustness and generalization compared to MLP in cross-subject scenarios. This observation can be attributed to the low-dimensional and low-frequency characteristics of the preprocessed insole pressure signals, where load-related variations are approximately linear or weakly nonlinear. However, SVR proves to be the most optimal choice among the evaluated regressors. While Elastic Net provides efficient linear regularized estimation, SVR’s ability to map input features into a high-dimensional space via kernel functions allows it to better accommodate the complex, non-linear biomechanical variations of the foot under load. This results in a more stable prediction profile, particularly at the boundaries of the tested load range. In contrast, MLP tends to amplify inter-subject variability, leading to higher prediction variance and a greater density of outliers in certain subjects and load ranges. For practical deployment in industrial exoskeletons, managing these outliers is a critical safety requirement; an erroneous torque output triggered by an extreme prediction error could compromise user stability. Therefore, the inherent robustness of SVR makes it a more reliable candidate for real-time control logic.

The evaluation on unseen load levels further highlights the practical relevance of the proposed method. In real industrial settings, exoskeleton users frequently encounter continuous and previously unknown load magnitudes rather than a small set of predefined weights. The ability of SVR and Elastic Net to infer excluded load levels (3 kg, 6 kg, and 9 kg) with acceptable accuracy indicates that the models capture underlying load-dependent pressure redistribution mechanisms instead of merely memorizing discrete load labels. This property is essential for adaptive gravity compensation and impedance modulation, where smooth interpolation across load levels is required to avoid abrupt or unsafe assistance behavior.

The comparison between channel-based and map-based representations reveals complementary strengths. Channel-based regression demonstrates strong generalization and computational efficiency, making it well-suited for real-time applications with limited onboard resources. In contrast, the map-based approach with full fine-tuning (Full-FT) exhibits superior stability under high-load conditions, such as the 9 kg level. This performance reversal suggests that as load magnitude increases, the biomechanical response—including the shift of the Center of Pressure (CoP) and the expansion of the foot arch—creates complex, non-linear spatial patterns that are better captured by deep convolutional architectures than by discrete channel vectors. These findings indicate that the choice of representation should be guided by specific application requirements, balancing computational constraints against the need for high-load precision.

Several limitations of this study should be acknowledged. First, the current experiments focus on quasi-static lifting tasks and do not explicitly model temporal dynamics, which may play a more significant role during highly dynamic movements. Second, the map-based approach was evaluated only in an intra-subject setting, and its cross-subject generalization remains to be validated. Finally, the proposed load estimation framework has not yet been integrated into a closed-loop exoskeleton control system. Future work will address these limitations by incorporating temporal modeling, expanding the subject pool, and evaluating the impact of load estimation accuracy on adaptive assistance performance in real-time control scenarios.

\vspace{1em}
\section{CONCLUSIONS}
This study demonstrated the feasibility of estimating variable lifting loads in industrial upper-limb exoskeletons using insole pressure sensors and machine learning. Experimental results across five subjects showed that channel-based regression, particularly SVR, achieves robust performance and generalizes to unseen load levels, supporting continuous load estimation for adaptive assistance. A preliminary evaluation of map-based representations further indicated potential advantages under high-load conditions. Overall, the findings validate insole pressure sensing as a practical modality for real-time load estimation in exoskeleton applications, with future work focusing on temporal modeling and closed-loop control integration.

\bibliographystyle{IEEEtran}
\bibliography{mylib}

\end{document}